# AN EXTENDED WEIGHTED PARTITIONING AROUND CLUSTER HEAD MECHANISM FOR AD HOC NETWORK

S.Thirumurugan[1] and E. George Dharma Prakash Raj[2]

[1]Department of Computer Applications, JJ College of Engineering & Technology, Trichirapalli, Tamilnadu, India
*s.thiru.gan@gmail.com*
[2]Department of Computer Science and Engineering, Bharathidasan University, Trichirapalli, Tamilnadu, India
*georgeprakashraj@yahoo.com*

***ABSTRACT***

*The wireless network places vital role in the present day communication scenario. The ad hoc nature of wireless communication adds flavour to suit various real world applications. This improves the performance of the network tremendously while the clustering mechanism gets added to the ad hoc network. It has been found out that the existing WCA lacks in forming efficient clusters. Thus, this work proposes an Extended weighted partitioning around cluster head mechanism by considering W-PAC as a base to form clusters. The cluster members are configured with IPv6 address. This IPv6 clusters formed through W-PAC will be taken further for validation to determine the perfectness of clusters. The cluster formation and maintenance have been implemented in C++ as a programming language. The cluster validation has been carried out using OMNET++ simulator.*

## 1. INTRODUCTION

The ad hoc networks are contemporary networks which are playing vital role in this dynamic scenario. It is possible to extract maximum fruitfulness of the networks on real world applications by applying the proper mechanisms on ad hoc scenario networks.

The network faces downside in performance while it grows continuously. There are various parameters like bandwidth, energy level of node, neighbors of nodes and traffic decides the efficiency of the network. These parameters are affected due to constant growth of the network.

To provide solution to this ever growing unprecedented network the clustering mechanism has been brought up as a supportive technique to provide stability to the network. This mechanism reduces the routing table entry and reduces the burden at the node levels. This also adds up the power conservation strategy. Under clustering the change happens in topology can be handled by cluster maintenance strategy. Since the nodes are tied up under clusters the control packets exchange happens within the boundary of the cluster. This leads to reduction in bandwidth consumption.

In ad hoc scenario the movement of the node is not compulsory, depends on the type of application. While the node moves the topology of the network changes. This has to be informed to all the nodes in the network over the wireless links. This task consumes high bandwidth. The traffic increases and bandwidth consumption rises while the number of nodes increases. Since the increase of nodes will proportionately increase the control packets exchange among the nodes.

Having understood the wireless networks critical role in present situation, the network with an improved performance has been crucial since the scalability will limit the functionality. In order to put the efficiency value of the ad hoc network to the high state multiple parameters are being well thought-out. It is obviously realized that the distance parameter alone will not be able to fix on the efficiency of clustering mechanism. Since, The nodes are in wireless network are facing energy drain as a problem with respect to time. Furthermore, the cluster head needs maximum energy among the nodes in a cluster to act as a transceiver. The cluster head election based on multiple parameters has formed an efficient clusters. These clusters need to be validated further to certify the perfectness of the clusters. Thus this study puts the focus on validating the clusters formed using weighted partitioning around cluster head procedure.

This paper has been orchestrated as follows. Section1 puts down the introductory part. Section2 discusses related work. Section3 tells about existing weighted clustering technique. Section4 describes the proposed clustering and validation technique. Section5 shows the experimental results and analysis. Section6 speaks about the scope of this work. Section7 concludes this research work.

## 2. RELATED WORKS

The role of clustering in ad hoc networks has been realized when AODV[1] incorporates the clustering mechanism. This added mechanism enhances the functionality of the network.

The application of the clustering mechanism[2] with AODV as a routing protocol in the real world scenario has been indispensible. This shows that clustering technique makes the network to be suitable for various applications. The clustering mechanism PAC[3] over the k-means approach tells the purpose of parameters in cluster formation. These parameters decide the efficiency level of clusters. This study also confirmed that k-means takes more time when the number of nodes are high in count. This work lacks in implementation and also the sample set of nodes are small in size.

The PAC has shown good results when the number of nodes are less. It leaves many nodes as non clustered nodes. The Ex-PAC [4] came out as an extension to PAC which takes entire nodes and produces the maximum clusters. The cluster formation process ultimately improved in Ex-PAC procedure. This approach concludes that Ex-PAC has outperformed k-means in terms of computational speed.

The cluster formation time has been significantly reduced in W-PAC[5] procedure. In this approach the clusters are created using multiple parameters. These clusters may or may not be considered as perfect in their existence.

The clustering can also be formed based on the signal strength[6] between the cluster head and nodes belong to cluster. The cluster head has been computed using the signal strength expression. The cluster head will select the nodes for the cluster on the basis of signal strength.

The dominating sets are identified in the clustered network. In which the minimum independent set[7] can be constructed and the tree structure of the same can be formed later. The connected dominating set algorithm has been a backbone to form the clusters.

The cluster head functionality can be tampered by the malicious node which behaves like cluster head. The SWCA[8] proposed a secured weighted clustering algorithm to keep the network away from such malicious nodes.

The NWCA[9] has been proposed to improve the weight based algorithms through changing methodology of parameter calculation for weight. The degree computation has been changed to mean connectivity degree. This novel method also considers the energy level of the nodes to play the role of cluster head.

The DWCA[10] considers the cluster formation based on weight, mobility factor and cluster maintenance. The new node addition to cluster has been handled through distinct approach by this protocol.

The re-clustering should be based on identifying the strength of the existing clusters. The role of various indices[11] on evaluating the cluster should be understood very well. The cluster

classification[[12] also plays key role in determining the perfectness of the clusters. Those classifications are of numeric, discrete and partitioned types. It also finds out the preferred clustering method for a given sample set of nodes.

The cluster formation algorithms devised so far lacks in obtaining an efficient cluster in terms of time, maintenance and validation to tune-up the clustered networks.

## 3. WCA

The existing weighted clustering algorithm clearly tells that the distance parameter alone will not be sufficient to choose the cluster head. It is essential to compute the weight based on degree of node, energy level, distance of node from their neighbors and mobility speed of the node.

$$W(Ni) = q1*Deg(Ni) + q2*M(Ni) + q3*D(Ni) + q4*E(Ni) \quad (1)$$

The weight has been computed based on above mentioned expression (1). The node with least weight value will be considered as cluster head at time period T. When cluster head has been exhausted with the energy level while communicating with the neighbor nodes or cluster heads, the cluster head re-election need to be handled. The cluster formation considers Euclidean distance as one of the parameter to compute the weight. This may take more time to form the clusters. This proportionately increases to the number of clusters belong to the network.

## 4. CLUSTER FORMATION & VALIDATION

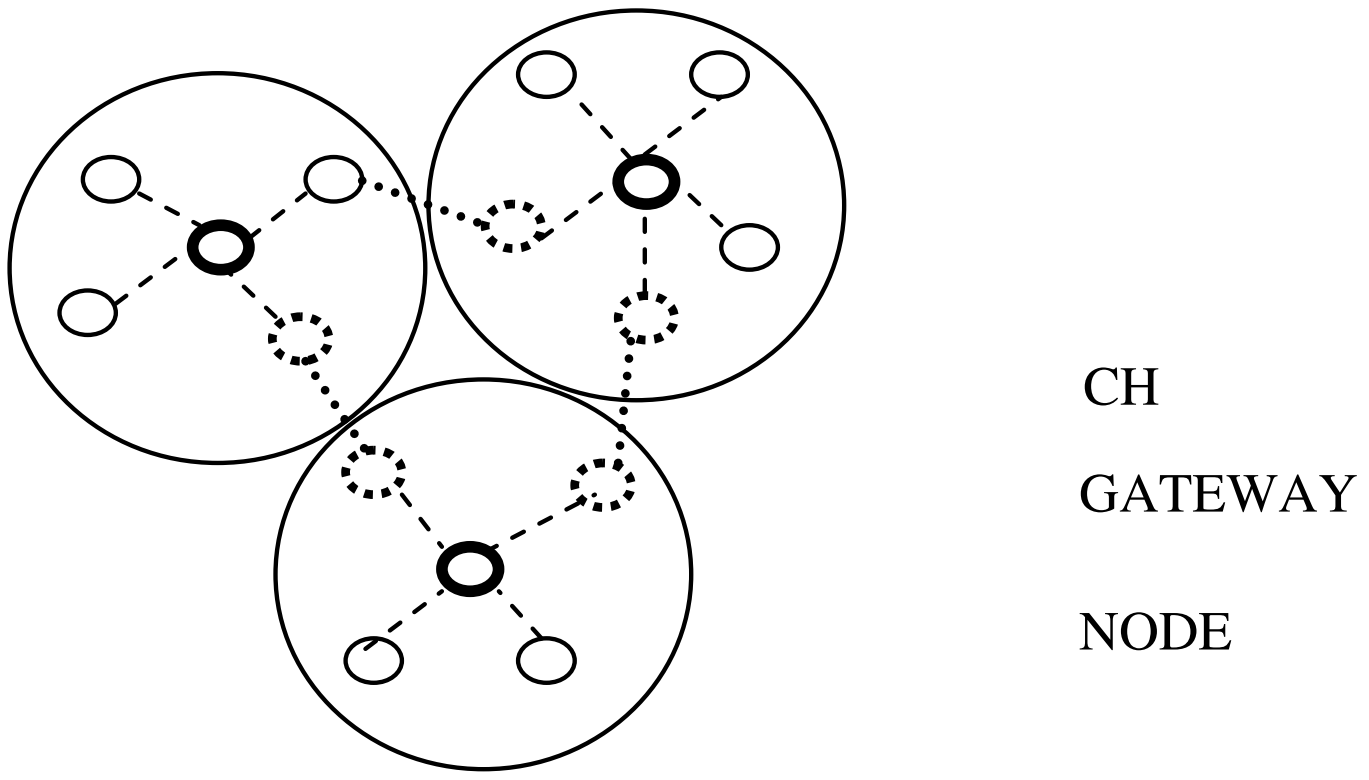

Figure 1. Clustered Network

The Figure.1 shows the clustered network structure. In this network each cluster has cluster head and gateway as a node to support inter cluster communication. The WCA algorithm faces high cluster formation time which leads to down side. In order to overcome this draw back the distance parameter has been replaced by Manhattan distance under W-PAC procedure. Since the distance parameter places major role in time consumption as far as the node weight computation is concerned. The following extended weight partitioning around cluster head procedure has been used to create cluster, elect cluster head, cluster maintenance and validate the clusters.

***Extended weighted Partitioning Around Cluster Head Procedure***

{
*Cluster Creation*
*Cluster Head Election*
*Cluster Maintenance*
*{*
*Mobility*
*Load Distribution*
*Cluster Head Rotation*
*Gateway Election*
*}*
*IPv6 Configuration*
*Cluster Validation*
}

## 4.1 Cluster Creation

(1) Initialize set of nodes as M.
(2) Compute the degree of node Ni.
(3) Deg (Ni) = 0.
(4) If ( i not equal to j)
(5) j = 1.
Add node Ni to Cluster Cm after computing node degree.
(6) Repeat the step 5 until j = M.

## 4.2 Cluster Head Election

(1) Create Clusters using W-PAC cluster creation.
(2) Cluster = Ci, P = Number of nodes in Ci.
(3) j = 1; Ni = $(U_t, V_t)$; Nj = $(U_{t-1}, V_{t-1})$;
(4) If ( i not equal to j)
If ( Manhattan Dist(Ni,Nj) < Radious )
Compute the Mobility speed of Node Ni belongs to Ci.
Compute the Distance between Ni and Nj.

(5) Repeat the step 4 until j = P.
(6) Assume the Energy of nodes E(Ni) for all the nodes.
(7) The weight of node Ni computed as follows,
W(Ni) = q1*Deg(Ni) + q2*M(Ni) + q3*D(Ni) + q4*E(Ni)
(8) Repeat the step 7 for all nodes belong to Ci.
(9) CHk = Min { $W(N_1), W(N_2), W(N_3) \ldots W(N_M)$ }.
(10) Repeat the step 2 through 9 for i = 1…..no of clusters.

The W-PAC procedure applies the weight computation on nodes. The Manhattan distance reduces the computation time due to the simplicity in calculation. This places very important role in determining the efficiency of the cluster formation procedure. The nodes which are falling within the radio range are added to Ci where i = 1…number of clusters and remaining nodes are included to NCn (Noncluster contains Nodes fall out of the radio range). The weight value will be calculated for the nodes which are falling within the specific radio range. The node with minimum weight will be elected as cluster head. This procedure is repeated for all the clusters.

## 4.3 Cluster Maintenance

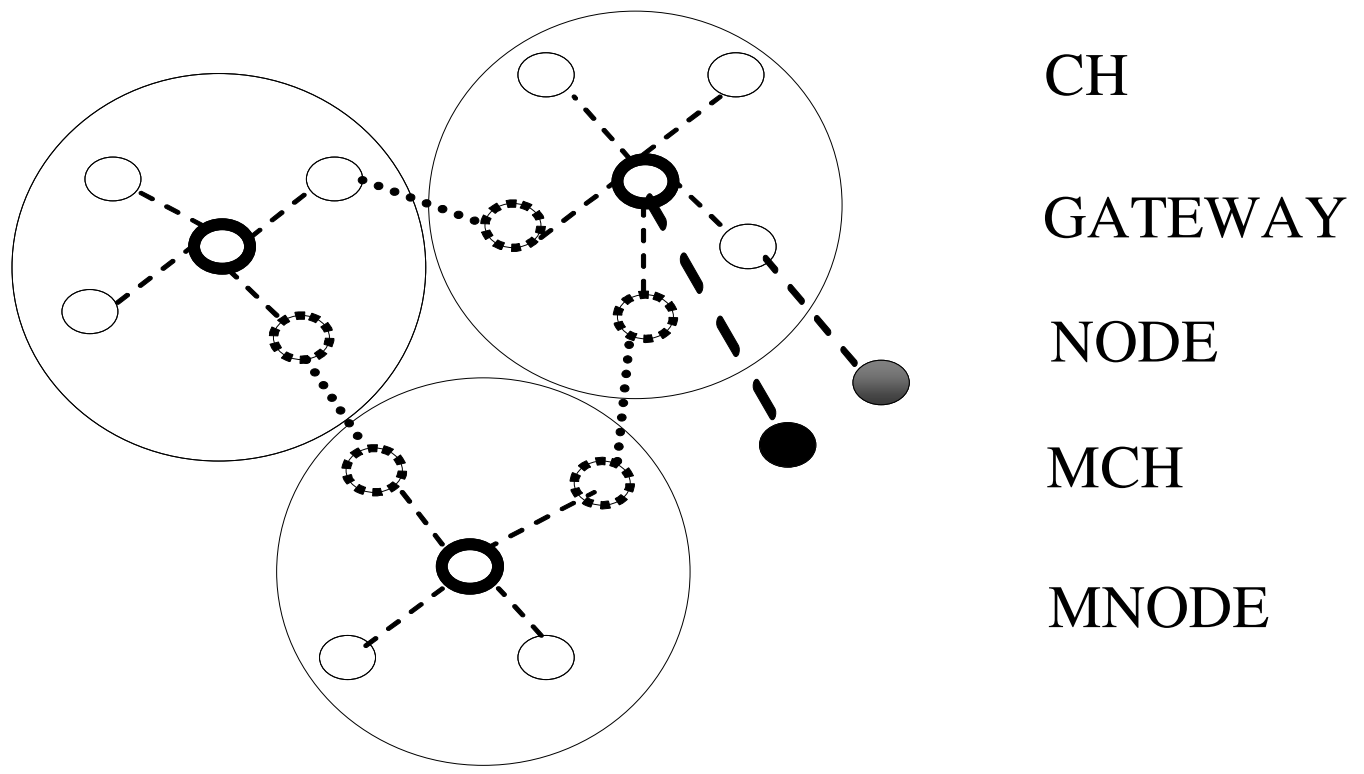

Figure 2. Clustered Network with Mobility

The Figure.2 shows the scenario where the node or the cluster head moves. This mobility property needs to be handled as part of maintenance. The W-PAC mechanism concentrates on local minimum of clusters. It means the node with minimum weight value within the cluster will become cluster head. In case of cluster head mobility(MCH) the re-election will happen within the cluster than finding the global minimum. When the node moves (MNODE) from a cluster to some other cluster then this will re-affiliate with the new cluster head of the cluster. The neighboring clusters when receives some new nodes then the re-clustering may happen while the new node weight is less than the weight of the previous cluster head. This will happen to the cluster locally. Since the cluster change has been confined with the cluster boundary. The reformation time of cluster will be less while this has been compared with the process of finding globally minimum weight value of the node as cluster head. This approach mitigates the overhead involved in cluster maintenance.

The cluster maintenance algorithm performs various tasks. Those tasks are shown as separate algorithms. These maintenance activities are performed local to the clusters. The re-clustering process by taking the entire network will not happen in this W-PAC procedure. This makes a major difference between WCA and W-PAC.

### 4.3.1 Algorithm:Mobility

(1) Input the Clusters created using Cluster creation and Cluster head Election procedure.
 Input : C1, C2…….Cn.
(2) Compute Mobility of Node MNi
 $MNi = | xi - xj | + | yi - yj|$

 If (MNi not equal ‘0’) then
  Mobility of Node MNi = True.
 Else
  Mobility of Node MNi = False.
(3) If (MNi = ‘True’)
 If (Node Ni =’Ordinary Node’)
  If ( Ci contains Ni)
   If (weight of Ni > weight of CHi)

Ni gets affiliated under CHi
Else
If (Node Ni = 'Clusterhead')
Begin
Re-elect the CH of Cluster from which the CH has moved.
Affiliate the mobile CH as new 'Ordinary Node' to neighbor Cluster.
End

(4) Repeat step 3 for each MNi = True.

This algorithm shows the steps to be followed to maintain the cluster when an ordinary node moves or cluster head moves across the clusters.

#### 4.3.2 Algorithm:Load Distribution

(1) Input : CHi and Ci.
(2) If energy of CHi < threshold then
Re-elect CHi of the Ci.
Make the new CHi to act as Proxy CHi.
(3) Repeat the step 2 for all Ci.

This algorithm has steps to be followed to ensure the load distribution at the cluster head level. The proxy CHs are selected when the cluster heads are exhausting their energy level within the cluster.

#### 4.3.3 Algorithm:CH Rotation

(1) Input : CHi and Ci.
(2) If energy of CHi < threshold and Time >Ti then
Present CHi resigns.
Re-elect new CHi of the Ci.
(3) Repeat the step 2 for all Ci.

The CH Rotation algorithm makes the cluster head role to be rotated among the member nodes of cluster based on the energy level and elapsed time as cluster head.

#### 4.3.4 Algorithm:Gateway Election

(1) If Ni ∈ (Ci, Cj) then
Gateway = Command Gateway.
(2) For each Ni in Ci do
// distributed Gateways
Gateway of Ci = Max(Distance(Ni, CHi))
(3) List of Gateways of Clusters Ci.

The Gateway election algorithm specifies the steps to elect common or distributed gateways.

### 4.4 IPv6 Configuration

1) Input the Clusters Created using Cluster Creation and Cluster Head procedure.
2) Assume CH(Cluster head) node as DHCPv6 server.
3) for each cluster Ci
Begin

Consider each node within Ci as DHCPv6 Client.
The Client sends hello message to DHCPv6 server.
The DHCP Server Configures the DHCPv6 Client Node.
End

4) Repeat the step 3 for i = 1..... Number of clusters.

The functionality of this network has been similar to IPv4 network. The cluster head runs DHCPv6 server. This server configures the member node which runs client process. The clients are assigned with IPv6 address as part of configuration.

### 4.5 Cluster Validation

The IPv6 address configured clusters can be validated to identify the perfectness of the clusters. The validation indices are used to make this study possible. The cluster formation has shown good results in wireless networks. But this clusters formed using the W-PAC procedure may or may not be perfect in nature. There must some techniques to identify the clusters quality. This quality may signify the compactness of clusters and physical proximity of the clusters. The validation procedure has been one such method to do the earlier mentioned technique. It is a function of ratio of sum of within cluster distribute to between-cluster disconnection. The computation of one of the cluster belong to a pair as follows,

$$\left\{ \begin{array}{ll} T1 = \sum_{i=1}^{n} (Ni - CH1) \ , & CH1 \in Cn \\ Cn(Ni) = T1/n \ , & n = 1 \ldots \text{number of nodes} \end{array} \right\}$$

The computation of another cluster belong to the pair of clusters as follows,

$$\left\{ \begin{array}{ll} T2 = \sum_{i=1}^{n} (Nj - CH2) \ , & CH2 \in Cn \\ Cn(Nj) = T2/n \ , & n = 1 \ldots\ldots \text{number of nodes} \end{array} \right\}$$

The DB index can be calculated as follows,

$$\frac{1}{n} \sum_{i=1}^{n} \max_{i \neq j} \left\{ \frac{Cn(Ni) + Cn(Nj)}{C(Ni, Nj)} \right\} \qquad (2)$$

In the aforementioned formula (2) , Cn(Ni) finds out the average distance of all nodes from Cluster head in a cluster.Cn(Nj) also points out the same as Cn(Nj). The value of 'n' indicates the cluster number. C(Ni,Nj) represents the distance between two cluster heads belongs two different clusters.

$$\left\{\begin{array}{ll} 0 < DB < 1 & i = 1,2 \ldots n \\ \text{DB} > 0.5 & \textit{Clusters are less} \text{ compact and closer to each other} \\ \text{DB} < 0.5 & \textit{Clusters are compact} \text{ and are far from each other} \end{array}\right\}$$

*Validation Procedure :*

1. Create the clusters using the Ex-PAC procedure.
2. Calculate Cn(Ni).
3. Repeat step 2 for i = 1 to number of nodes.
3. Calculate Cn(Nj).
4. Repeat step 3 for j = 1 to number of nodes.
5. Compute Cn(Ni,Nj).
6. Find Max(Cn(Ni) + Cn(Nj) / Cn(Ni,Nj)).
6. Compute DB index for each cluster pair.
7. Repeat the steps 2 through 4 for each pair of clusters in a Network.
8. if (DB < 0.5) then " Clusters are compact
   Else
   If (DB > 0.5) then "Clusters are less compact"

The above algorithm identifies the compactness of clusters.

## 5. EXPERIMENTAL RESULTS

This work has been implemented in C++ as a programming language and the results are tabulated. The validation has been carried out using OMNET++ simulator. It has been carried out with the system configuration of 32 bit AMD processor, 2GB RAM and windows XP as an operation system. The simulation has been done for 25 nodes and 50 nodes. The size of the network has been 40 × 80 meters square space. The weight values used are w1=0.7, w2=0.2, w3=0.05, w4=0.05. The Table.1 shows simulation parameters. In this study the transmission range has been fixed to 20m distance for the sample set of nodes 25 as well as 50. These parameters with the specified values created two clusters for sample set of nodes considered. The energy threshold decides the time to distribute the cluster head load among the nodes.

Table 1. Simulation Parameters

| Parameter | Values |
|---|---|
| N (Number of Nodes) | 25 , 50 |
| Space (area) | 40 × 80 |
| Tr (Transmission range) | 20m |
| Simulation Time | 5 sec |
| Weighing Factors (w1,w2,w3,w4) | 0.7,0.2, 0.05,0.05 |
| Threshold(Energy) in Units | 500 |

### 5.1 Analysis of Cluster Formation Time

The weight has been calculated for each node of sample set. The node with least weight has been chosen as cluster head. In the case of 25 nodes two clusters have been formed with the radius value 20. The node6 and node19 have been the cluster heads of cluster1 and cluster2 respectively. Similarly for 50 nodes as the sample size, node 6 and node 40 have been the cluster heads of cluster1 and cluster2 respectively.

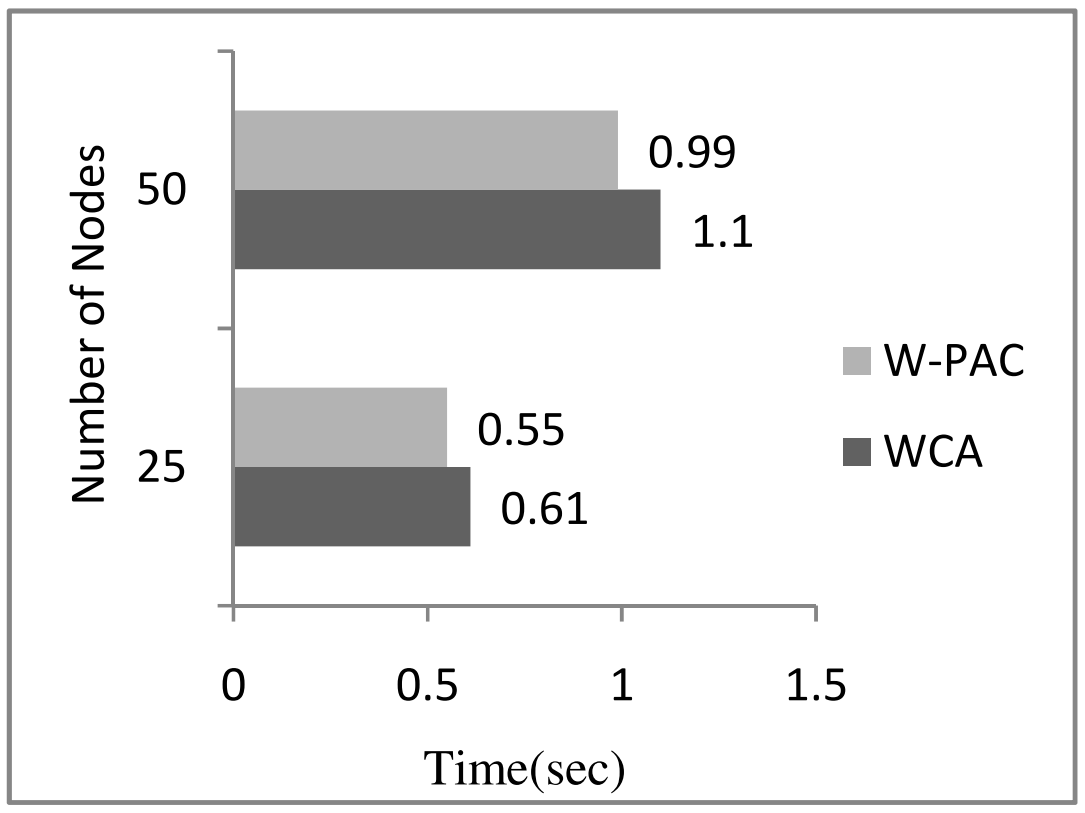

Figure 3. WCA Vs W-PAC

The graphical illustration shows the execution time of W-PAC and WCA. The Figure.3 clearly reveals that cluster formation time of W-PAC has been significantly lower than WCA at time T1.

## 5.2 Analysis of Cluster head Mobility

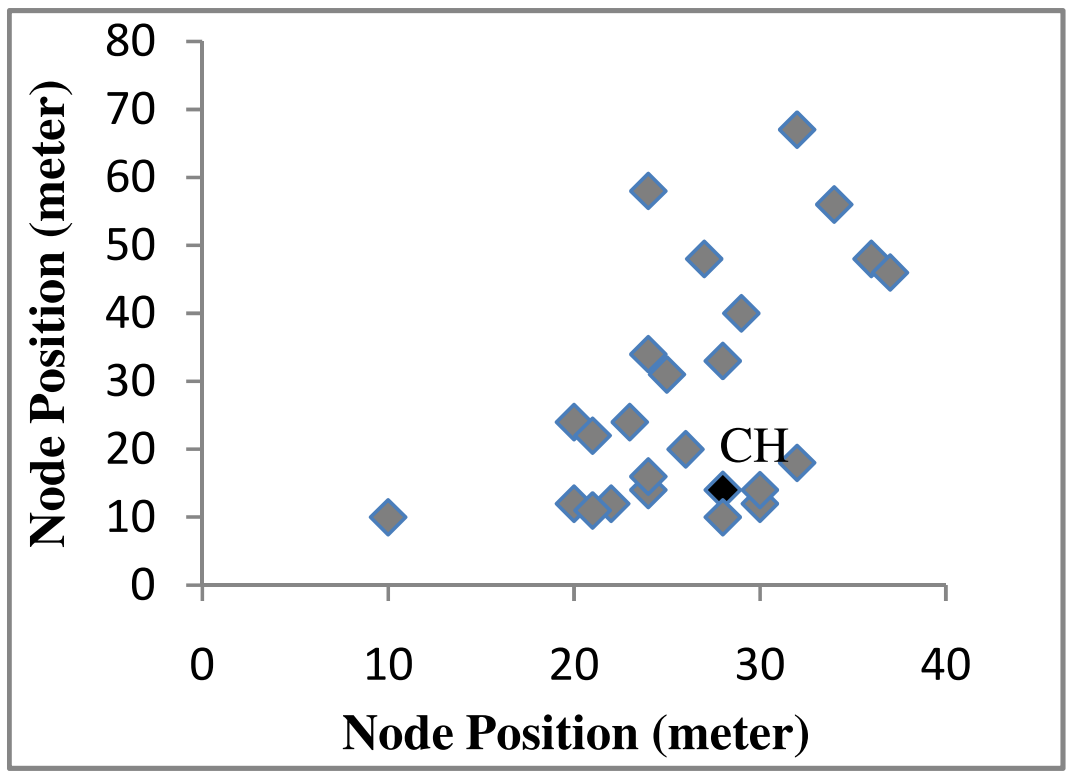

Figure 4. Result before Cluster head's Mobility

The Figure.4 shows the result of cluster1 at time T1 for 25 nodes as a sample set. The node[6] has been elected as the cluster head. Similarly for 50 nodes the clusters have been formed and the cluster heads have been identified based on the weight value of the nodes. The node6 position at time T1 has been [28,14]. Due to mobility nature of cluster head the position has been changed from [28,14] to [38,34]. As a result of this the cluster head gets changed from node6 to node12 which resides at [32,18]. This has been clearly shown in Figure.5. The old cluster head reduced the intra cluster relationship strength and new cluster head has got elected. This scenario has been followed under cluster maintenance phase when cluster head makes movement. The nodes movement without disturbing the existing clusters and their heads makes the re-affiliation under new cluster head.

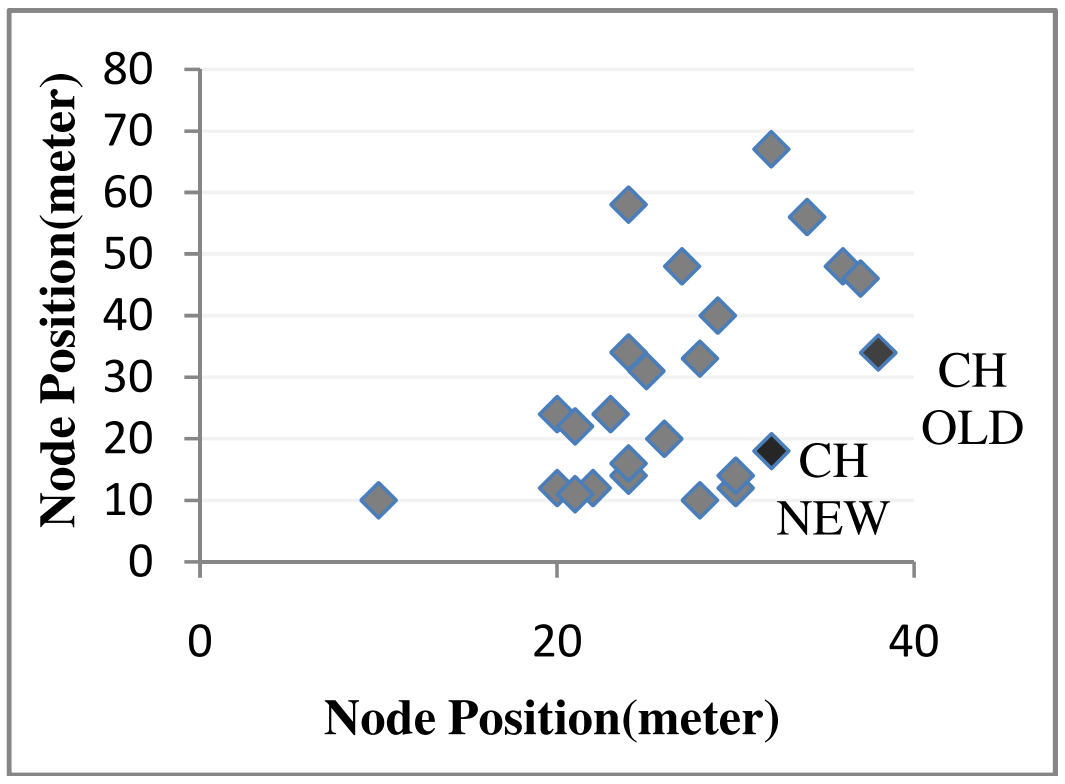

Figure 5. Result after Cluster head's Mobility

## 5.3 Analysis of Cluster head Rotation

The cluster head rotation is based on energy level and elapsed time the node acted as cluster head. The cluster heads are decided based on weight values computed earlier. These weight values have to be sorted in ascending order where the lowest value would be the first cluster head of the respective cluster. If any node has moved in the mean time those nodes need to be excluded.

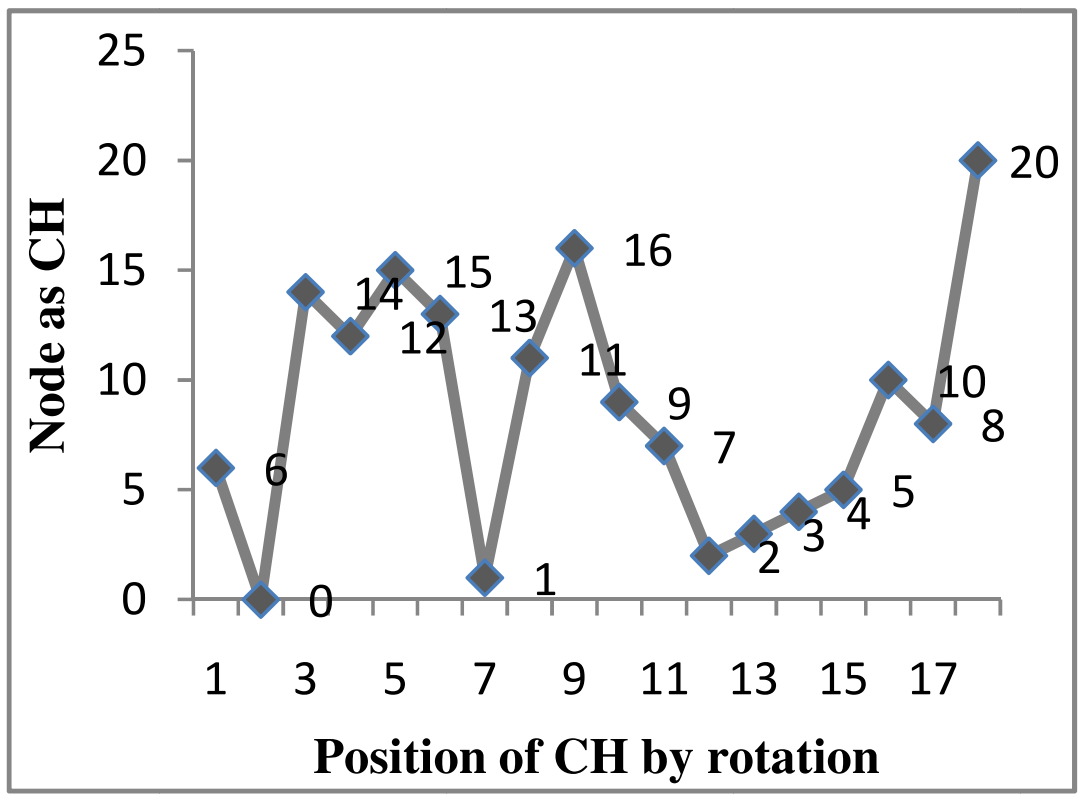

Figure 6.  Result of Cluster head Rotation in Cluster C1

The Figure.6 shows the cluster heads order to be followed within the cluster C1. This comes with the assumption that the mobility of the node will reduce the hope of a node becoming cluster head of that cluster.

### 5.4 Validation of Clusters

The validation has been carried on clusters formed using W-PAC for the sample set of 25 nodes and 50 nodes. The index computation has been implemented using OMNET++ as a simulator.

Table. II Results of Cluster Validation at T1

| Nodes | Clusters | DB Index |
|---|---|---|
| 25 | C1 & C2 | 0.52 |
| 50 | C1 & C2 | 0.429 |

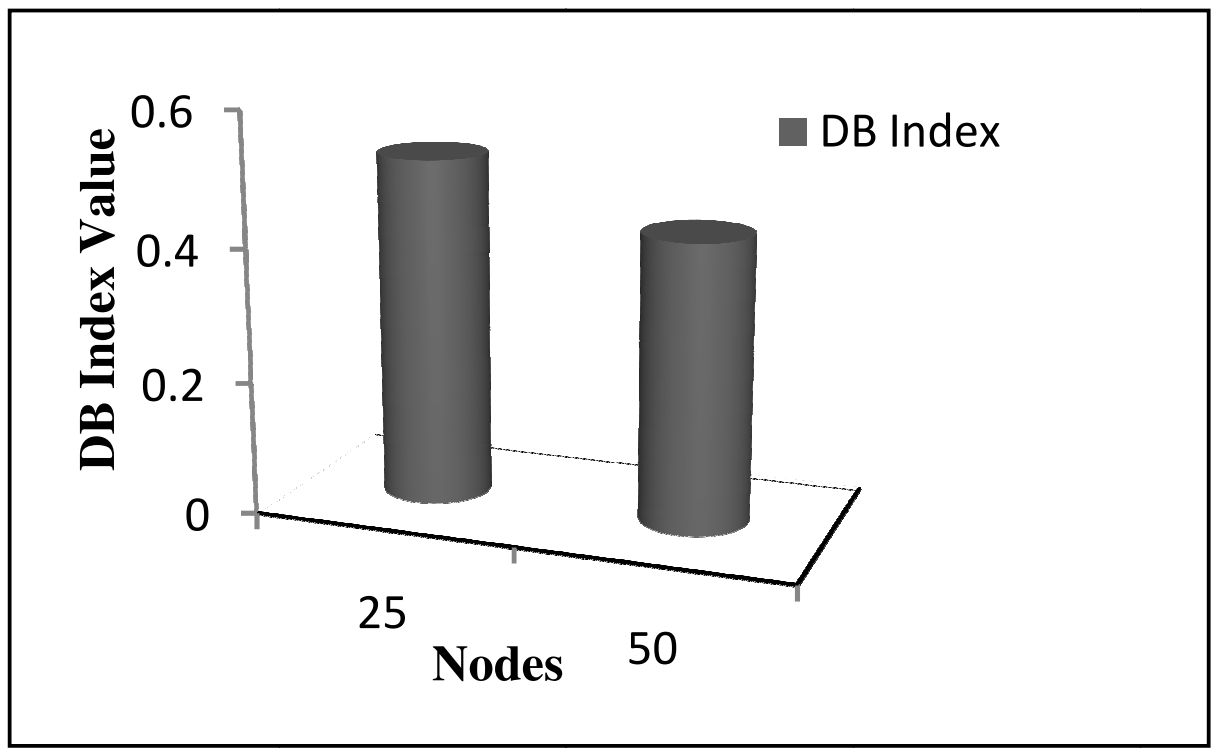

Figure 7.  Graphical Validation Result

The Table.II shows that number of sample set of nodes and corresponding to that computed DB index value. This also shows that two clusters are formed for 25 and 50 nodes. The Fig.7 graphically illustrates DB index value for the sample set of nodes considered for analysis. The computed value obviously reveals that the compactness of the clusters has been less while 25 nodes are considered. This parameter is slightly high for 50 nodes when compared with 25 nodes. Thus, the low value of DB index indicates that the clusters overlapping possibility has been very low. The sample set of 50 nodes says that the nodes belong to the set tightly coupled with the cluster head pertaining to the cluster.

## 6. FUTURE DIRECTIONS

This work should address the issues when IPv4 and IPv6 nodes get into single cluster. The validation should also consider various indices to identify the perfect and valid clusters.

## 7. CONCLUSION

This study identifies the benefits of W-PAC procedure over WCA procedure. The extended weighted partitioning around cluster head mechanism adds cluster validation procedure to the existing W-PAC procedure. This addition proves the perfectness of the clusters formed using W-PAC procedure. This study also helps to find out the re-clustering time period.

**Authors**

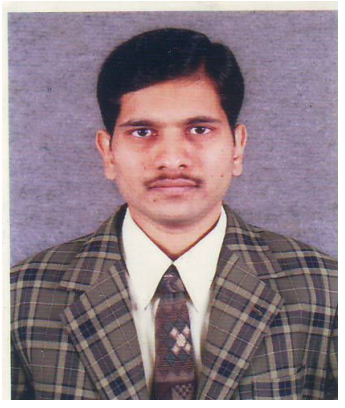

**Mr.S.Thirumurugan** completed his Masters Degree in Computer Applications and Master of Philosophy in Computer Science. He has around 9yrs of experience in teaching field which includes his association with the research work. He has published his work in international journals, presented two papers at the international level and also two papers at the national level. His area of research work falls on wireless networks and their applications on real world scenario.

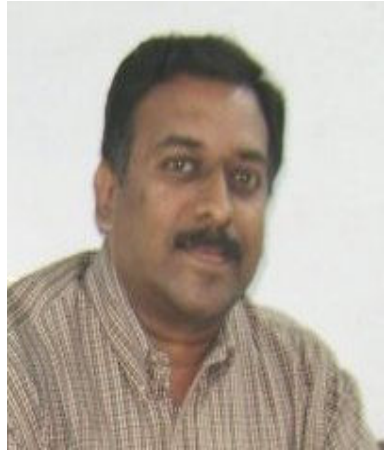

**Dr.E. George Dharma Prakash Raj** completed his Masters Degree in Computer Science and Master of Philosophy in Computer Science in the years 1990 and 1998. He has also completed his Doctorate in Computer Science in the year 2008. He has around twenty-one years of Academic experience and thirteen years of Research experience in the field of Computer Science. Currently he is working as an Asst.Professor in the Department of Computer Science and Engineering at Bharathidasan University, Trichy, India. He has published several papers in International Journals and Conferences related to Computer Science and has been an Editorial Board Member, Reviewer and International Programme Committee Member in many International Journals and Conferences.